\documentclass{article}
\usepackage{amsmath, amssymb, graphics}

\newcommand{\mathsym}[1]{{}}
\newcommand{\unicode}{{}}

\title{\noindent Taxon Size Distribution in a Time-Homogeneous Birth and Death Process.}
\author{\begin{tabular}{ll}
Panagis Moschopoulos  &  Max Shpak\\
Dept. of Mathematical Sciences  &  Dept. of Biological Sciences\\
Univ. of Texas at El Paso &   Univ. of Texas at El Paso\\
El Paso TX, 79968   &El Paso TX, 79968\\
pmoschopoulos@utep.edu &mshpak@utep.edu\\
\end{tabular}}
\date{\today}
\begin{document}
\maketitle
\begin{abstract}

The number of extant individuals within a lineage, as exemplified by counts of species numbers across genera in a higher taxonomic category,
is known to be a highly skewed distribution. Because the sublineages (such as genera in a clade) themselves follow a random birth process, deriving the distribution
of lineage sizes involves averaging the solutions to a birth and death process over the distribution of time intervals separating the origin of the
lineages. In this paper, we show that the resulting distributions can be represented by hypergeometric functions of the second kind. We also provide
approximations of these distributions up to the second order, and compare these results to the asymptotic distributions and numerical approximations
used in previous studies. For two limiting cases, one with a relatively high rate of lineage origin, one with a low rate, the cumulative probability
densities and percentiles are compared to show that the approximations are robust over a wide rane of parameters. It is proposed that the probability density distributions of lineage size may have a number of relevant applications to biological
problems such as the coalescence of genetic lineages and in predicting the number of species in living and extinct higher taxa, as these
systems are special instances of the underlying process analyzed in this paper.
\end{abstract}

\subsubsection*{Running  Heading} Family Size Distribution

\subsubsection*{Keywords} \textit{birth-death process, taxon size, hypergeometric function, phylogeny, genealogy}

\pagebreak

\section*{1. Introduction}

The number of individuals in a genetic lineage (family size) and the formally equivalent problem of the number of species within
genera or other higher taxa has been of wide interest to demographers and biologists alike (e.g. Watson and Galton 1875, Yule, 1924, Anderson 1974,
Burlando 1990). Empirically, the distributions of family sizes in humans and taxon sizes in a number of organisms have been documented to be consistent with the distributions predicted by simple stochastic models.

The most straightforward approximation to the distribution of lineage or taxon sizes is the pure birth process, which entails
a certain probability of an individual giving birth or one species giving rise to another per unit time, ignoring death or extinction. In its details, this is clearly
an unrealistic model, but it does provide a reasonable first order approximation for the diversification of higher taxa during their early history,
when extinction is infrequent, or in lineages (such as small populations of bacteria growing on a rich medium) where the death rates are very small
compared to birth rates.

A basic model (Yule 1924, Feller 1950, Bailey 1968) represents a pure birth process with birth probability $\lambda $ per unit time. In this model,
the instantaneous rate of change in the probability of observing n individuals is given by the following differential equation

$$ \text{        }\frac{dp_n}{dt}=\lambda (n-1)p_{n-1}-\lambda  n p_n\text{                                         }. \eqno(1.1)$$

Under the initial condition \(p_1\)(0)=1 (a single ancestor for the lineage at time zero) and 0 for all $n>$1, this differential equation has the solution

$$\text{        }p_n(t)=e^{\lambda t}(1-e^{\lambda t})^{n-1}\text{                          }.\eqno(1.2)$$

More realistically, if there is an intrinsic death rate (or, for species and higher taxa, extinction rate)  $\mu $, then one has a time-homogeneous
birth and death process defined by the differential equation, e.g. Kendall (1948), Darwin (1956), Keiding (1975),

$$\text{        }\frac{dp_n}{\text{dt}}=\lambda (n-1)p_{n-1}-(\lambda +\mu ) n p_n+\mu (n+1)p_{n+1}\text{    }\text{ for } n>0\text{       }$$
$$\frac{dp_0}{dt}=\mu p_1\text{                      }.\eqno(1.3)$$
The solutions for $\lambda \neq \mu $ given \(p_1\)(0)=1 are given by

$$\text{        }p_n(t)=\frac{(\lambda -\mu )^2\ \, \text{exp}[-(\lambda -\mu )t]}{(\lambda -\mu \, \text{exp}[-(\lambda -\mu )t])^2}\left(\frac{\lambda
-\lambda \, \text{exp}[-(\lambda -\mu )t]}{\lambda -\mu  \,\text{exp}[-(\lambda -\mu )t]}\right)^{n-1}\text{  }\text{ for } n >0\text{        }$$

$$p_0(t)=\frac{\mu
-\mu \, \text{exp}[-(\lambda -\mu )t]}{\lambda -\mu \, \text{exp}[-(\lambda -\mu )t]}\text{                                      }.\eqno(1.4)$$
In the special case where $\lambda $=$\mu $, the solutions simplify to:

$$\text{        }p_n(t)=\frac{(\text{$\lambda $t})^{n-1}}{(\text{$\lambda $t}+1)^{n+1}}\text{ for } n >0\text{        }$$
$$p_0(t)=\frac{\text{$\lambda
$t}}{\text{$\lambda $t}+1}\text{ .                                                               }\eqno(1.5)$$

For the purposes of many empirical studies, in which one does not know the precise number of once-extant lineages which are currently
extinct (i.e. $n=$0 individuals in the lineage or species in the clade at time $t$, with an indeterminate number in previous time intervals), it is useful
to work with the truncated distributions describing the probability densities for strictly positive values of n, i.e.

\[\text{        }P_n(t)=\text{  }\frac{p_n(t) }{1- p_0(t)}\text{for                                  }n\geq 1\]
which, using the expressions in (1.4), evaluates to

$$\text{        }P_n(t)=\frac{(\lambda -\mu )\,\text{exp}[-(\lambda -\mu )t]}{\lambda -\mu  \,\text{exp}[-(\lambda -\mu )t]}\left(\frac{\lambda -\lambda
 \,\text{exp}[-(\lambda -\mu )t]}{\lambda -\mu  \,\text{exp}[-(\lambda -\mu )t]}\right)^{n-1}\text{  }\text{ for } n \geq 1\text{     }\eqno(1.6)$$
while for $\lambda $=$\mu $,

$$\text{        }P_n(t)=\frac{(\lambda t)^{n-1}}{(\lambda t+1)^n}\text{  }\text{ for } n \geq 1\text{.     }\eqno(1.7)$$

Strictly speaking, in order to compare the distributions of family sizes to those predicted in (1.4) or (1.6), or to infer birth and
death rates using maximum likelihood estimators, the lineage ages must be known. This requirement becomes particularly problematic when one
compares the number of individuals in sublineages, for example the number of species in different genera within a taxonomic family or order. Because the sublineages
arose at different times (i.e. obviously, not all genera within a clade originated at the same moment), comparisons of the number of species in a
genus or any other subclade must be weighted by the differences in subclade ages.

When the age of a lineage is known, as is the case for phylogenetic trees with branch lengths that have been callibrated using fossil data and $\texttt{"}$molecular
clocks$\texttt{"}$ (sensu Zuckerkandl and Pauling 1965, Kimura 1980), the number of individuals or species in each lineage can be compared to the
predictions of equation (1.4) by applying the appropriate time values. For phylogenies with known branch lengths and resolved split times, Harvey et al (1994) and Nee et al (1994,
1995) presented a method whereby the origination and extinction rates $\lambda $ and $\mu $ could be estimated using maximum likelihood methods.
A similar approach, not requiring conditioning on the age of the first split, is given by Rannala (1997) and in Felsenstein (2004,
pgs. 564-9).

Such detailed data is not reliably available for the majority of phylogenetic trees or gene genealogies, and when it is, the confidence intervals on the age estimates can be very wide. To address the problem of rate estimation when molecular clock estimates are absent or unreliable,
Reed and Hughes (2002) derived distributions of genera sizes weighted by the time intervals separating generic branching events. Under the assumption
that within a lineage (clade), subclades such as genera within families arise at a rate $\rho $, the authors used an exponential approximation for
the distribution of the generic ages. This representation is not limited to any taxonomic group, as it applies to any lineage in which constituent
sublineages are defined by some unique characters, genetic markers, or in the case of humans, family names.

The time at which a genus within a clade of age $\tau $ arises is given by

$$\text{        }f(t)=\frac{\rho  e^{-\text{$\rho $t}}}{1-e^{-\rho \tau }}\text{                            }\eqno(1.8)$$
which, for $\tau $ $\to \infty $, gives

$$\text{        }f(t)=\rho  e^{-\text{$\rho $t}}.\text{                               }\eqno(1.9)$$
Using (1.9) to weight the probability distributions \(p_n\) at time t, Reed and Hughes defined

$$\text{        }q_n(\tau)=\int _0^{\tau }\rho \, p_n(t)\,e^{-\rho t}dt\text{ .         }\eqno(1.10)$$
Reed and Hughes derived a generating function and an algorithm for calculation of \(q_n\)($\tau$) recursively from a limiting case (when $n$$\to \infty
$). It will be shown that given certain choices of changes in variables, a closed form expression for this integral can be evaluated, and a number
of useful approximations to the \(q_n\)($\tau$) distribution can be made, of which the asymptotic approximation calculated by Reed and Hughes is an important
limiting case.

\section*{2. Results}

\subsection*{Pure Birth Process}

We will begin with the distribution for a pure birth process, which is a reasonable approximation when $\lambda \gg \mu $.

$$\text{        }q_n(\tau )=\int _0^{\tau }\rho \, \,exp[-(\lambda +\rho )t](1-exp[-\lambda t])^{n-1}dt\text{ .                }\eqno(2.1)$$
Under the assumption that the lineage is ancient relative to the age of any sublineage, we take the limit as $\tau $ $\to $ $\infty $. Then, applying a change
of variables\\ $y=e^{\lambda t}$, the resulting integral is

$$\text{        }q_n= \rho  \int _0^1 y^{\rho /\lambda }(1-y)^{n-1}dy\text{        }\eqno(2.2)$$
which, for n$\geq $1, evaluates to

$$\text{        }q_n= \text{Beta}(\rho /\lambda +1,n)=\frac{\Gamma \left(\frac{\rho }{\lambda }+1\right)\Gamma (n)}{\Gamma \left(n+\frac{\rho }{\lambda
}+1\right)}\text{                           }.\eqno(2.3)$$
Using an asymptotic expansion of the gamma ratios involving n above (e.g. Exton, 1978, Luke 1969), we get the following: This can be further simplified to an expansion of the ratios of gamma functions, up to the order $n^{-2}$,

\begin{eqnarray*}
\hspace{-.75in} q_n = \frac{\Gamma (\rho /\lambda +1)} {n^{\rho /\lambda +1}}
\left\{1 -\frac{\frac{\rho }{\lambda }(1+\frac{\rho }{\lambda })}{2n}\right. \\
&\left. \hspace{-1in} +\frac{\left(1+\frac{\rho}{\lambda }\right)\left(2+\frac{\rho }{\lambda }\right)}{24n^2}\left[3\left(\frac{\rho }{\lambda }+1\right)^2+\frac{\rho }{\lambda }\right]+O(n^{-3})\right\}.
\end{eqnarray*}
It follows from the above that, as $n$$\to \infty $, the probability can be approximated by

$$\text{         }q_n = \Gamma (\rho /\lambda +1) n^{-\rho /\lambda -1}.\eqno(2.4)$$

\subsection*{Birth and Death Processes}

As in the analyses of Reed and Hughes (2002), we consider three different birth and death models. The first, with $\lambda >\mu $
(birth rate higher than death rate), is perhaps the most biologically significant, because this is the only birth and death process in which a lineage
has a nonzero probability of persisting indefinitely.

We will evaluate the function \(Q_n\) for the truncated solutions to the birth and death process given by equation (1.6), because most
of the relevant empirical data involves $n>$0. For convenience we use the parameters $\omega $=$\lambda $$-$$\mu $ and $\theta
$=$\mu $/$\lambda $, so that when the birth rate is greater than the death rate, $\omega >$0 and $\theta <$1. The integral over the truncated distribution can be written as

\[\text{        }Q_n=\int _0^{\infty }\frac{\rho \omega  \,\exp[-(\omega +\rho )t]}{1-\theta  \,\exp[\omega t]}\left(\frac{1- \,exp[-\omega t]}{1-\theta  \,\exp[-\omega t]}\right)^{n-1}dt.\]
Applying the changes of variable

\[\text{          }y(t)=\frac{1- exp[-\omega t]}{1-\theta  \,exp[-\omega t]},\; \text{dt}=\frac{(1-\theta )\,\text{dy}}{\omega (1-y)(1-\theta y)}\]
and noting that the integral limits are $y(0)=$ 0 and $y$($\infty $)=1, we obtain

$$\text{        }Q_n=\rho (1-\theta )\int _0^1y^{n-1}(1-y)^{\rho /\omega }(1-\text{$\theta $y})^{-\rho /\omega -1}dy\text{.                  }\eqno(2.5)$$
If we do the same for Equation (1.4) by integrating over the part of the distribution where $n$$\geq $1, an expression of similar form is derived

$$\text{        }q_n=\frac{\rho (1-\theta )}{\omega }\int _0^1y^{n-1}(1-y)^{\rho /\omega }(1-\text{$\theta $y})^{-\rho /\omega }dy\text{
          }.\eqno(2.6)$$
The two integrals given above evaluate to constant multiples of a hypergeometric function (Exton 1978, Luke 1969), which has the integral representation

$$\text{      }\int _0^1y^{a-1}(1-y)^{c-a-1}(1-\text{$\theta $y})^{-b}dy\text{  }=\text{  }\frac{\Gamma (c-a)\Gamma (a)}{\Gamma (c)}\, _2F_1(a,b,c,\theta
)\text{      }\eqno(2.7)$$
and \(\, _2\)\(F_1\) is the Gauss hypergeometric function that is defined by the series

$$\text{      }\, _2F_1(a,b,c,\theta )=\sum _{k=0}^{\infty } \frac{(a)_k(b)_k}{(c)_k}\theta ^k  \eqno(2.7a)$$
where

\[\text{     }(a)_k=a(a+1)\text{...}(a+k-1)=\frac{\Gamma (a+k)}{\Gamma (a)}.\]
(Exton 1978, Luke 1969). In equations (2.5) and (2.6), we have $a=n$ and $b$=1+$\rho $/$\omega $. In the first instance, $c=n$+$\rho $/$\omega $+1, and $c=n$+$\rho
$/$\omega $+2 in the second. Using (2.7) and (2.7a) in (2.5) we obtain

$$Q_{n}=\text{  }\rho (1-\theta ) \Gamma \left(1+\frac{\rho }{\omega }\right) A\sum _{k=0}^{\infty } \left(1+\frac{\rho }{\omega
}\right)_k B \frac{\theta ^k}{k!}\text{.  }\eqno(2.8)$$
The coefficients $A=A(\rho,\omega,n)$ and $B=B(\rho,\omega,k,n)$ are given by

\[A(\rho,\omega,n)=\frac{\Gamma (n)}{\Gamma \left(n+\frac{\rho }{\omega }+1\right)},\]

\[B(\rho,\omega,k,n)=\frac{(n)_k}{\left(n+\rho/\omega+1\right)_k}=\frac{\Gamma(n+k)\, \Gamma(n +\rho/\omega +1)}{\Gamma(n) \,\Gamma( n+k+\rho/\omega+1)}.\]

It is well known that the ratios of Gamma functions can be expanded in terms of 1/$n$, see for example Luke (1969). The expansion is in terms of Bernoulli polynomials, see also Moschopoulos and Mudholkar (1983) for similar asymptotic expansions of hypergeometric functions. The approximations to follow are lengthy and rather tedious, but nevertheless straightforward.

Second order approximations to the terms A and B are given by:

\[A= n^{-\left(1+\frac{\rho }{\omega }\right)} \left(1-\frac{\frac{\rho }{\omega } \left(1+\frac{\rho }{\omega }\right)}{2 n
}+\frac{\left(1+\frac{\rho }{\omega }\right) \left(2+\frac{\rho }{\omega }\right) \left(3 \left(\frac{\rho }{\omega }\right)^2+\frac{\rho }{\omega
}\right)}{24 n^2 }\right)\times\left(1+O(n^{-3})\right)\]
To solve for B, we use the following:

\[\frac{\Gamma (n+k)}{\Gamma (n)}= n^k\left(1+\frac{k (k-1)}{2 n}+\frac{k(k-1)\text{  }\left(3 (k-1)^2-k-1\right)}{24 n^2})\right)\times\left(1+O(n^{-3})\right)\]

\[\frac{\Gamma
\left(n+\frac{\rho }{\omega } +k+1\right)}{\Gamma \left(n+\frac{\rho }{\omega } +1\right)}= n^k\left(1+\frac{k \left(k+\frac{2 \rho }{\omega
}+1\right)}{2 n}+\frac{k(k-1) \left(3 \left(k+\frac{2 \rho }{\omega }+1\right)^2-k-1\right)}{24 n^2}\right)\]

\noindent $\times\left(1+O(n^{-3})\right)$\\

\[\frac{\Gamma
\left(n+\frac{\rho }{\omega } +k+1\right)}{\Gamma \left(n+\frac{\rho }{\omega } +1\right)}= \]

\[n^k\left(1+\frac{k \left(k+\frac{2 \rho }{\omega
}+1\right)}{2 n}+\frac{k(k-1) \left(3 \left(k+\frac{2 \rho }{\omega }+1\right)^2-k-1\right)}{24 n^2}\right)\times\left(1+O(n^{-3})\right)\]
which allows one to rewrite B as a series expansion:

\[B= \left({1+\frac{k (k-1)}{2 n}+\frac{k(k-1)\text{  }\left(3 (k-1)^2-k-1\right)}{24 n^2}}\right)\]

\[
\times \left
(1+\frac{k(k+\frac{2\rho}{\omega}+1}{2n}+\frac{k(k-1)(3(k+\frac{2\rho}{\omega}+1)^2-k-1)}{24n^2}
\right)^{-1}\times\left(1+O(n^{-3})\right)\]

\[=1-\frac{k}{n}\left(1+\frac{\rho
}{\omega }\right)
+\frac{1}{n^2}\phi(\rho,\omega,k)+O(n^{-3})\]

\noindent where
$$\phi(\rho,\omega,k)=\left(k^2\left(\frac{\rho ^2}{2 \omega ^2}+\frac{3 \rho }{2 \omega }+1\right)-k \left(\frac{\rho ^2}{2 \omega ^2}+\frac{\rho
}{2 \omega }\right)\right).$$
Collecting terms, we obtain the following approximation up to order $n^{-2}$

$$Q_n= \rho (1-\theta )\Gamma \left(1+\frac{\rho }{\omega }\right)n^{-1-\frac{\rho }{\omega }}  \left[1-\frac{\frac{\rho
}{\omega } \left(1+\frac{\rho }{\omega }\right)}{2 n }
+\frac{\frac{\rho }{\omega } \left(1+\frac{\rho }{\omega }\right) \left(3 \left(\frac{\rho
}{\omega }\right)^2+\frac{\rho }{\omega }\right)}{24 n^2 }\right]$$
$$\times \sum _{k=0}^{\infty } \left(1+\frac{\rho }{\omega }\right)_k
\left[1-\frac{k}{n}\left(1+\frac{\rho
}{\omega }\right)
+\frac{\phi(\rho,\omega,k)}{n^2}\frac{\theta ^k}{k!}\right]\times\left(1+O(n^{-3})\right).\eqno(2.9)$$
The terms in the above approximation underscore the fact that the shape of the distribution \(Q_n\) is determined by the magnitude
of $\rho $/$\omega $, the ratio of the rates at which new lineages (genera) arise to the net birth rate of individuals or species. This is because
in the limit as $\tau \to \infty $, the absolute magnitudes of the two rate parameters do not contribute to the ratio of lineage number to lineage size. Basically, a small value of $\rho $/$\omega $ implies that over sufficient
time, there will be a comparatively low number of lineages rich in individuals, while a large value results in many lineages
with relatively few individuals.

It is also remarked that the series terms

\[\sum _{k=0}^{\infty } k\left(1+\frac{\rho }{\omega }\right)_k\frac{\theta ^k}{k!} \ ,\   \sum _{k=0}^{\infty } k^2\left(1+\frac{\rho }{\omega
}\right)_k\frac{\theta ^k}{k!}\]
have to be computed numerically, just as the hypergeometric function must be estimated using numerical methods. As there is no closed-form expression
for equation (2.9), the value of these approximations is to show the rate of convergence of first and second order terms (as functions of $n^{-1}$) to
the exact solution for different birth and death rates.

In contrast, a closed form expression does exist for the asymptotic limit, as $n\to \infty $. Unlike the terms involving higher powers of k in (2.9), the first series term within the parenthesis converges to

\[\unicode{ }\sum _{k=0}^{\infty } \left(1+\frac{\rho }{\omega }\right)_k\frac{\theta ^k}{k!}=(1-\theta )^{-1-\rho /\omega }\text{.}\]
By ignoring all factors of higher order than $n^{-1}$, we have

$$\text{        }Q_n\approx \rho (1-\theta )^{1-\rho /\omega }\Gamma \left(1+\frac{\rho }{\omega }\right) n^{-\rho /\omega -1}\text{
                                  }\eqno(2.10)$$
as in Reed and Hughes (2002), apart from the constant factor of 1/$\omega $ for the truncated distribution. It is noteworthy that in equation (2.10) the term $\theta
$ only appears as a constant factor, while in (2.10) its magnitude determines the rate at which the series term converges. The efficacy of both the asymptotic and $n^{-2}$ approximations will be investigated numerically in the next section, where they are compared to the exact hypergeometric function solutions.

We next direct our attention to the case of a "declining" linage where $\lambda <\mu $, i.e. when the birth or speciation rate is less than the death or extinction rate. In this scenario, the parameter values $\omega<$0 and $\theta <$1. Using the same changes of variables as in the derivation of equation (2.6), the integral defining the probability density is

$$\text{        }Q_n=\rho (1-\theta )\int _0^{1/\theta }y^{n-1}(1-y)^{\rho /\omega }(1-\text{$\theta $y})^{-\rho /\omega -1}dy\text{
 .     }\eqno(2.11)$$
With a further change of variables, z=$\theta $y, we again derive an expression with integral limits of 0 and 1,

$$\text{        }Q_n=\rho (1-\theta )\theta ^{1-n}\int _0^1z^{n-1}\left(1-\frac{z}{\theta }\right)^{\rho /\omega }(1-z)^{-\rho /\omega -1}dz\text{
  .                }\eqno(2.12)$$
The solution to this integral is a hypergeometric function of a form similar to that of equation (2.6). Here, the parameters are a=n, b=-$\rho $/$\omega $,
and c= n-$\rho $/$\omega $+1. In place of { }$\theta $ in the series expansion, we have 1/$\theta $, so that (2.11) ultimately evaluates to

$$\text{        }Q_n=\rho (1-\theta )\theta ^{1-n}\text{  }\Gamma \left(-\frac{\rho }{\omega }\right)\text{  }A\sum _{k=0}^{\infty } \left(-\frac{\rho
}{\omega }\right)_kB\frac{\theta ^k}{k!}\text{ .             }\eqno(2.13)$$
Note that because $\omega <$0, $- \rho $/$\omega >$0. For the above distribution, the coefficients A and B are

\[\unicode{ }A=\frac{\Gamma (n)}{\Gamma \left(n-\frac{\rho }{\omega }\right)}\ ,\  B(\rho,\omega,k,n)=\frac{(n)_k}{\left(n-\frac{\rho }{\omega }+1\right)_k}=\frac{\Gamma(n+k)\, \Gamma(n -\frac{\rho}{\omega } +1)}{\Gamma(n) \,\Gamma( n+k-\frac{\rho}{\omega }+1)}.\]

An approximation in terms of leading orders of 1/n is possible for (2.13) using the same approach as was used in (2.9) i.e. by rewriting the Gamma functions
as truncated polynomial ratios.

For the A, the second order approximation is:

\[\unicode{ }A = n^{\frac{\rho }{\omega }} \left(1-\frac{\frac{\rho }{\omega } \left(\frac{\rho }{\omega }+1\right)}{2 n }+\frac{\left(\frac{\rho
}{\omega }-1\right)\left(\frac{\rho }{\omega }-2\right) \left(3 \left(1+\frac{\rho }{\omega }\right)^2-\frac{\rho }{\omega }-1\right)}{24 n^2}\right)\times\left(1+O(n^{-3})\right)\]

To evaluate B, we have the same ratio $\Gamma $(n+k)/$\Gamma $(n) in the numerator as in the case of $\lambda > \mu$, while the denominator term $\Gamma $( 1-$\rho/\omega$+k)/ $\Gamma$(n-$\rho/\omega$) is:

\[\unicode{ }n^k\left(1+\frac{k \left(k-\frac{2
\rho }{\omega }\right)}{2 n}+\frac{(k-1) k \left(3 \left(k-\frac{2 \rho }{\omega }-1\right)^2-k-1\right)}{24 n^2}\right)\times\left(1+O(n^{-3})\right)\]
Using the series expansion of coefficients A and B, we obtain the following second order approximation for equation (2.13),

\[\hspace{-2.5in}Q_n = \rho (1-\theta )\theta ^{1-n}\text{  }\Gamma \left(-\frac{\rho }{\omega }\right) n^{\frac{\rho }{\omega }}\]
\[ \hspace{-.5in}\times \left\{1-\frac{\frac{\rho
}{\omega } \left(\frac{\rho }{\omega }+1\right)}{2 n }+\frac{\left(\frac{\rho }{\omega }-1\right)\left(\frac{\rho }{\omega }-2\right) \left[3 \left(1+\frac{\rho
}{\omega }\right)^2-\frac{\rho }{\omega }-1\right]}{24 n^2}\right\}\]
$$\times \sum _{k=0}^{\infty } \left(-\frac{\rho }{\omega }\right)_k\left[1+\frac{k}{n}\left(\frac{\rho
}{\omega }+1\right)+\frac{\phi(\rho,\omega,k)}{n^2}\left(\frac{\theta ^k}{k!}\right)\right]+O(n^{-3}).\eqno(2.14) $$
Recall that $\omega <$0, so that the coefficients -$\rho $/$\omega $ are positive.

The asymptotic approximation to this distribution, calculated for $n\to \infty $, is given by the following equation:

\[\text{        }Q_n\approx \rho \,\theta ^n(1-\theta )^{1-\rho /\omega }\Gamma \left(1+\frac{\rho }{\omega }\right) n^{-\rho /\omega -1}\text{
.                                         }\eqno(2.15)\]
The special case where $\lambda $=$\mu $ (exactly zero net growth rate) is not likely to occur in biological systems, therefore we do not analyze this scenario in detail. In those rare instances where net growth rate is close to zero, the distributions can be approximated by using the equations for the cases where growth rate is positive and negative and taking limits as $\mu $ approaches $\lambda $.

\subsection*{A Numerical Assessment of the Approximations}

The accuracy of the second order (2.9) and asymptotic approximations (2.10) are assessed by comparing their probability densities to those of the exact solutions (2.8) for a range of parameters values. The examples considered here are the biologically most relevant case where the
birth rate is higher than the death rate, such that $\omega >$0 and $\theta <$1.

Attention is focused on two parameters: $\rho $/$\omega $ and $\theta $. The first is the ratio of the rate at which new lineages arise
to the net birth rate of individuals or species within that lineage. As was discussed above in connection with the distribution (2.9), the shape of the probability
density function depends principally on the ratio $\rho $/$\omega $. When $\rho $/$\omega \ll $1, the distribution will be characterized by very few
lineages with large numbers of individuals, while as $\rho $/$\omega \to $1, the distribution will be skewed to the left due to the presence of many
lineages with few (perhaps only a single) individual. The accuracy of the approximation is examined for different values of this ratio, so that
in one instance $\rho $ and $\omega $ are of approximately the same magnitude, while in another $\rho $ and $\omega $ differ by an order of magnitude.

It is expected from the summation term in equation (2.9) that the second order approximation should be most accurate when $\theta
\ll$1, corresponding to a low birth rate relative to the death rate, because the series converges rapidly when $\theta $ is near zero.
To compare the accuracy of the approximations as a function of the rate parameters, we will consider cases where of $\theta $=0.01, 0.1, and 0.4.
The predictions on the accuracy of the respective approximations are given qualitative support by Figures 1 and 2, which plot the cumulative density functions

$$\sum_{N=1}^n Q_N$$

\noindent for the exact solution \(Q_n\) given by equation (2.8), the order \(n^{-2}\) approximation of equation (2.9), and the asymptotics in equation (2.10). In the first
set of figures, the rates at which new lineages (e.g. $\texttt{"}$genera$\texttt{"}$) are born is of the same order of magnitude as the net birth
rate, i.e. $\rho $=0.02, $\omega $=0.05, while in the second set, there is a tenfold difference between the birth rates of lineages and individuals, with $\rho $=0.01 and $\omega
$=0.1.\\

\pmb{\text{INSERT} \text{ FIGURES } 1a-c\text{  }\text{ and } 2a-c \text{  }\text{ HERE}}\\\

The relationship between the size of $\theta $ and the accuracy of approximation (2.9) can be seen by comparing Figures 1a and 2a
with 1c and 2c. In Figures 1a and 2a, $\theta $=0.01, so that the second order approximation performs better than the asymptotic approximation. The exact
solutions and second order approximations start to diverge when $\theta $=0.1, both in Figures 1b and 2b. For the case where $\theta $=0.4, the second order
approximation is quite poor, being consistently outperformed by the asymptotic approximation. This is because equation (2.10), unlike (2.9), has a closed form expression
and does not itself have to be estimated numerically.

As predicted, the asymptotic approximation begins to break down when the net birth rate (of species or individuals) as $\rho$, the rate at which new sublineages (family groups in a genealogy or genera in a clade) arise, approaches the value of $\omega $, the birth or speciation rate. This is made apparent by comparing the difference
between the exact solutions and the asymptotic approximations in Figures 1a-c (where $\rho $ is of the same magnitude as $\omega $) with those shown Figures
2a-c, where $\rho $ is an order of magnitude smaller than $\omega $. The greater deviation between the asymptotic and the exact solutions is due
to the fact that when $\rho $/$\omega $ is reasonably large, the coefficients of the $n^{-1}$ and $n^{-2}$ terms cannot be ignored, except when n is
very large (of the order of $\sim $\(10^3\) or greater), so that the second order approximation can match the exact distribution quite closely given parameter values where the asymptotic approximation is inaccurate.

A more precise assessment of the approximation accuracies can be made by comparing cumulative distributions functions. These are given in Table 1 for n=10, 100, 200, 500, 1000, and 10000 and are compared to exact and asymptotic values for the same set of birth and death rates that were shown in
Figures 1-2.\\

\pagebreak

Table 1. Prob.[$N(t)\leq n \mid \frac{\rho}{\omega}, \theta$]

\vskip.3in
\(
\begin{array}{cccccccccc}
  {\rho /\omega } &  {\theta } &  {P(10)} &  {P(50)} &  {P(100)} &  {P(500)} &  {P(1000)} &  {P(2000)} &  {P(10000)} &
 {\text{Dist}} \\
  {\text{\textit{$0.4$}}} &  {\text{\textit{$0.01$}}} & 0.580 & 0.780 & 0.834 & 0.915 & 0.937 & 0.953 & 0.977 & {\text{\textit{$E$}}} \\
  {\text{\textit{$0.4$}}} &  {\text{\textit{$0.01$}}} & 0.575 & 0.777 & 0.832 & 0.914 & 0.936 & 0.956 & 0.977 & {\text{\textit{$S$}}} \\
  {\text{\textit{$0.4$}}} &  {\text{\textit{$0.01$}}} & 0.604 & 0.794 & 0.845 & 0.920 & 0.941 & 0.956 & 0.979 & {\text{\textit{$A$}}} \\
  {\text{\textit{$0.4$}}} &  {\text{\textit{$0.1$}}} & 0.577 & 0.778 & 0.833 & 0.914 & 0.936 & 0.953 & 0.978 & {\text{\textit{$E$}}} \\
  {\text{\textit{$0.4$}}} &  {\text{\textit{$0.1$}}} & 0.576 & 0.777 & 0.832 & 0.914 & 0.936 & 0.952 & 0.977 & {\text{\textit{$S$}}} \\
  {\text{\textit{$0.4$}}} &  {\text{\textit{$0.1$}}} & 0.604 & 0.794 & 0.844 & 0.920 & 0.941 & 0.956 & 0.979 & {\text{\textit{$A$}}} \\
  {\text{\textit{$0.4$}}} &  {\text{\textit{$0.4$}}} & 0.563 & 0.770 & 0.826 & 0.911 & 0.933 & 0.951 & 0.976 & {\text{\textit{$E$}}} \\
  {\text{\textit{$0.4$}}} &  {\text{\textit{$0.4$}}} & 0.682 & 0.829 & 0.871 & 0.934 & 0.950 & 0.963 & 0.982 & {\text{\textit{$S$}}} \\
  {\text{\textit{$0.4$}}} &  {\text{\textit{$0.4$}}} & 0.604 & 0.794 & 0.845 & 0.920 & 0.941 & 0.956 & 0.979 & {\text{\textit{$A$}}} \\
  {\text{\textit{$0.1$}}} &  {\text{\textit{$0.01$}}} & 0.271 & 0.429 & 0.490 & 0.616 & 0.664 & 0.709 & 0.803 & {\text{\textit{$E$}}} \\
  {\text{\textit{$0.1$}}} &  {\text{\textit{$0.01$}}} & 0.270 & 0.428 & 0.489 & 0.615 & 0.664 & 0.709 & 0.802 & {\text{\textit{$S$}}} \\
  {\text{\textit{$0.1$}}} &  {\text{\textit{$0.01$}}} & 0.275 & 0.432 & 0.493 & 0.618 & 0.666 & 0.711 & 0.804 & {\text{\textit{$A$}}} \\
  {\text{\textit{$0.1$}}} &  {\text{\textit{$0.1$}}} & 0.270 & 0.429 & 0.489 & 0.615 & 0.664 & 0.709 & 0.802 & {\text{\textit{$E$}}} \\
  {\text{\textit{$0.1$}}} &  {\text{\textit{$0.1$}}} & 0.268 & 0.425 & 0.487 & 0.614 & 0.662 & 0.708 & 0.801 & {\text{\textit{$S$}}} \\
  {\text{\textit{$0.1$}}} &  {\text{\textit{$0.1$}}} & 0.275 & 0.432 & 0.493 & 0.618 & 0.666 & 0.711 & 0.804 & {\text{\textit{$A$}}} \\
  {\text{\textit{$0.1$}}} &  {\text{\textit{$0.4$}}} & 0.268 & 0.426 & 0.487 & 0.614 & 0.662 & 0.708 & 0.802 & {\text{\textit{$E$}}} \\
  {\text{\textit{$0.1$}}} &  {\text{\textit{$0.4$}}} & 0.300 & 0.448 & 0.506 & 0.628 & 0.674 & 0.718 & 0.810 & {\text{\textit{$S$}}} \\
  {\text{\textit{$0.1$}}} &  {\text{\textit{$0.4$}}} & 0.275 & 0.432 & 0.493 & 0.618 & 0.666 & 0.711 & 0.804 & {\text{\textit{$A$}}}
\end{array}
\)
\vskip.3in

For the smaller values of $\theta $, the cumulative probabilities of the second order approximation are very close to the exact distribution, while for $\theta $=0.5, the asymptotic is better. As was seen in the graphs,
the asymptotic cumulative probabilities diverge for $\rho $ of the same order of magnitude as $\omega $. The asymptotic approximation are independent of $\theta $, because this value only appears as a constant factor in equation (2.10), while it is a parameter in the hypergeometric function in (2.8) and the approximating sum in (2.9).

Another measurement of the approximation was based on a numerical calculation of percentiles (the calculations were implemented using
the \textit{ Mathematica} software package, with scripts available from the corresponding author by request) for each distribution, for p=0.05 and
p=0.01. The percentile values are shown in Table 2:\\

\pagebreak

Table 2. Upper 95th and 99th Percentiles Percentiles Given $\frac{\rho}{\omega}, \theta$

\vskip.3in
\(
\begin{array}{ccccc}
  % {\rho /\omega } &  {\theta } &  {p=0.05} &  {p=0.01} &  {\text{Dist}} \\
 {\rho /\omega } & {\theta } & {p=0.05} & {p=0.01} & {\mbox{Dist}} \\
   {\text{\textit{$0.4$}}} &  {\text{\textit{$0.01$}}} & 1743 & 49111 &  {\text{\textit{$E$}}} \\
  {\text{\textit{$0.4$}}} &  {\text{\textit{$0.01$}}} & 1798 & 50276 &  {\text{\textit{$S$}}} \\
  {\text{\textit{$0.4$}}} &  {\text{\textit{$0.01$}}} & 1491 & 43576 &  {\text{\textit{$A$}}} \\
  {\text{\textit{$0.4$}}} &  {\text{\textit{$0.1$}}} & 1778 & 49853 &  {\text{\textit{$E$}}} \\
  {\text{\textit{$0.4$}}} &  {\text{\textit{$0.1$}}} & 1801 & 50358 &  {\text{\textit{$S$}}} \\
  {\text{\textit{$0.4$}}} &  {\text{\textit{$0.1$}}} & 1491 & 43576 &  {\text{\textit{$A$}}} \\
  {\text{\textit{$0.4$}}} &  {\text{\textit{$0.4$}}} & 1949 & 53439 &  {\text{\textit{$E$}}} \\
  {\text{\textit{$0.4$}}} &  {\text{\textit{$0.4$}}} & 980 & 31356 &  {\text{\textit{$S$}}} \\
  {\text{\textit{$0.4$}}} &  {\text{\textit{$0.4$}}} & 1491 & 43576 &  {\text{\textit{$A$}}} \\
  {\text{\textit{$0.1$}}} &  {\text{\textit{$0.01$}}} & 251193 & 746770 &  {\text{\textit{$E$}}} \\
  {\text{\textit{$0.1$}}} &  {\text{\textit{$0.01$}}} & 251582 & 747932 &  {\text{\textit{$S$}}} \\
  {\text{\textit{$0.1$}}} &  {\text{\textit{$0.01$}}} & 249229 & 745463 &  {\text{\textit{$A$}}} \\
  {\text{\textit{$0.1$}}} &  {\text{\textit{$0.1$}}} & 251470 & 746954 &  {\text{\textit{$E$}}} \\
  {\text{\textit{$0.1$}}} &  {\text{\textit{$0.1$}}} & 252951 & 747932 &  {\text{\textit{$S$}}} \\
  {\text{\textit{$0.1$}}} &  {\text{\textit{$0.1$}}} & 249229 & 745463 &  {\text{\textit{$A$}}} \\
  {\text{\textit{$0.1$}}} &  {\text{\textit{$0.4$}}} & 252810 & 747839 &  {\text{\textit{$E$}}} \\
  {\text{\textit{$0.1$}}} &  {\text{\textit{$0.4$}}} & 241228 & 740048 &  {\text{\textit{$S$}}} \\
  {\text{\textit{$0.1$}}} &  {\text{\textit{$0.4$}}} & 249229 & 745463 &  {\text{\textit{$A$}}}
\end{array}
\)

\vskip.3in

As predicted from the previous results, it can be seen that the critical values \(n^*\) corresponding to each percentile are very similar for the
exact and second order approximate distributions when $\theta \ll$1. Similarly, the critical values \(n^*\) between the exact and second order diverge
for larger $\theta $, just as those of the asymptotic and exact distributions diverge for larger values $\rho $/$\omega $.

Taken together, these examples illustrate the efficacy of the approximations under a range of rate parameter values. It remains to be determined
whether the assumptions made in the derivations are realistic or sufficiently accurate for the biological processes being modeled.

\section*{3. Discussion}

The results derived in this paper give analytical predictions for the number of individuals per lineage in a distribution of lineages given a time-homogeneous birth and death process, weighted by a time-homogeneous birth rate for sublineages.
An obvious application of these distributions is to the problem that motivated this and similar earlier studies - the number of species within a genus or some higher taxonomic category. With modern techniques of phylogenetic inference, monophyletic subclades of any clade of interest can be identified (together with reasonable
bounds on age estimates in many instances) to yield an empirical frequency distribution of subclade sizes. These frequency distributions can then
be compared to the probability densities predicted from the exact solutions \(Q_n\).

It is of particular interest to look at $\texttt{"}$anomalously$\texttt{"}$
species rich subclades and evaluate the probability of their occurrence based on a null model, something which can readily be done by computing
the extremal probabilities. For example, if the average rate at which new genera arise in a sufficiently extant clade is 0.02 per million years,
the net speciation rate $\omega $=0.05, and the ratio of extinction to speciation rate is 0.1, then it can be seen from the appropriate entry in
Table 1 the probability of encountering a genus with over 10000 species is less than 0.025 according to the distributions derived under the assumptions of time homogeneity. If such unusually large genera are
encountered, that may indicate that the birth-death process driving the distribution is in fact not time-homogeneous.

Another application of these results to phylogenetic data is the estimation of speciation and extinction rates from taxon size distributions. The
most straightforward, and by far the most accurate, means of estimating these parameters is from direct paleontological data on origination and extinction rates, which is rarely available
at the level of detail required for likelihood based estimates. The approach of Harvey and colleagues (e.g. Harvey et al 1994, Nee et al 1994) does
not require fossil data, relying instead upon molecular clock estimates of subclade ages. In addition to the fact that the confidence intervals on
these estimators tend to be very large (Paradis, 2004), there is the fact that reliable clock data is not available for many taxa. Consequently,
the distributions in equations 2.8-2.10 can be used to give numerical estimates of the parameters, much as Reed and Hughes (2002) obtained recursively and in asymptotic approximations.

The underlying processes for which the distributions were derived are not limited to phylogenetic data on species abundance.
Indeed, any process generating new entities defined as lineages or families, that are themselves composed of individual
entities undergoing birth and death, can be analyzed using the models presented in this paper. For example, gene genealogies within populations and species often have constituent lineages
that are defined by a set of point mutations. This is of course the foundation of coalescent based approaches to population genetics, whereby
the history of lineages can be inferred from sequence data (e.g. Kingman 1982, Hudson 1991, Hein et al 2005, Wakeley 2008). The diversity of genotypes
and lineages is produced by a branching process where the birth and death rates can be parameterized by $\lambda $ and $\mu $, while the mutation
rate defining new lineages in infinite sites or infinite alleles models can be used as a measure of $\rho $. In this way, haplotype distributions
in population genetics can be analyzed in the same way as the taxonomic data described above, allowing one to determine whether the disproportionate representation of certain alleles or haplotypes is consistent with a random branching process, or whether the haplotype distribution must be explained by processes other than mutation and genetic drift.

The correspondence between the data and the analytical results in all of these cases will to a large part be limited by a number
of simplifying assumptions made in the derivations. In particular, the distributions \(Q_n\) were derived under the assumption that the time elapsed
since the initiation of the process is very great, essentially infinite. Therefore, for comparatively young lineages (young with respect to the birth
and death rates), one would expect the approximations to break down. These issues will be investigated using individual-based simulations and analysis of biological data in a forthcoming paper.

\section*{Acknowledgements}
Panagis Moschopoulos was supported by RCMI/NIH grant 5G12 RR008124 from the National Institutes of Health to the Border Biomedical Research Center (BBRC) at the University of Texas at El Paso (UTEP), and Max Shpak was supported by funding for research from the University of Texas at El Paso.

\section*{References}

Anderson, S. 1974. Patterns of faunal evolution. Quarterly Review of Biology, 49: 311-332\\
\\
Bailey, N.T.J. 1964. The elements of stochastic processes with applications to the natural sciences. John Wiley and Sons, New York.\\
\\
Burlando, B. 1990. The fractal dimension of taxonomic systems. Journal of Theoretical Biology 146: 99-114\\
\\
Darwin, J.H. 1954. The behavior of an estimator for a simple birth and death process. Biometrika 43: 23-31\\
\\
Exton, H. 1978. Handbook of hypergeometric integrals. John Wiley and Sons, New York.\\
\\
Feller, W. 1950. Probability theory and its applications. John Wiley and Sons, New York\\
\\
Felsenstein, J. 2004. Inferring phylogenies. Sinauer Associates, Sunderland.\\
\\
Harvey, P.H., R.M. May, and S. Nee 1994. Phylogenies without fossils. Evolution 48: 523-529\\
\\
Hein, J., M.H. Schierup, and C. Wiuf 2005. Gene genealogies, variation, and evolution: a primer in coalescent theory. Oxford University Press, Oxford.\\
\\
Hudson, R.R. 1991. Gene genealogies and the coalescent process. Oxford Surveys in Evolutionary Biology 7: 1-49\\
\\
Keiding, N. 1975. Maximum likelihood estimation in the birth and death process. The Annals of Statistics 3: 363-372\\
\\
Kendall, D.G. 1948. On the generalized birth and death process. Annals of Mathematical Statistics 19: 1-15\\
\\
Kimura, M. 1980. The neutral theory of molecular evolution. Cambridge University Press, Cambridge\\
\\
Kingman, J.F.C. 1982. The coalescent. Stochastic Processes and Applications 13: 235-248.\\
\\
Luke, Y.L. 1969. The special functions and their approximations. Academic Press, New York.\\
\\
Moschopoulos, P.G., and G.S. Mudholkar 1983. A likelihood ratio based normal approximation for the non-null distribution of the multiple correlation coefficient. Communications in Statistics: Simulation and Computation 12: 355-371\\
\\
Moschopoulos, P.G. 1985. The distribution of the sum of independent gamma random variables. Annals of the Institute of Statistics and Mathematics
37(A): 541-544\\
\\
Nee, S., R.M. May, and P.H. Harvey 1994b. The reconstructed evolutionary process. Philosophical Transactions of the the Royal Society of London B
344: 305-311.\\
\\
Nee, S., E.C. Holmes, R.M. May, and P.H. Harvey 1995. Estimating extinction rates from molecular phylogenies. Pgs 164-182 in J.H. Lawton and R.M. May, Extinction Rates. Oxford University Press, Oxford\\
\\
Paradis, E. 2004. Can extinction rates be estimated without fossils? Journal of Theoretical Biology 229: 1-30\\
\\
Rannala, B. 1997. Gene geneology in a population of variable size. Heredity 78: 417-423\\
\\
Reed, W.J. and B.D. Hughes 2002. On the size distribution of live genera. Journal of Theoretical Biology 217: 125-135\\
\\
Wakeley, J. 2008. Coalescent theory: an introduction. Roberts Publishing, Greenwood Village.\\
\\
Watson, H.W. and F. Galton 1875. On the probability of extinction of families. Journal of the Anthropological Institute of Great Britain 4: 138-144\\
\\
Yule, G.U. 1924. A mathematical theory of evolution based on the conclusions of Dr. J.C. Willis. F.R.S. Philosophical Transactions of the Royal Society of London B. 213: 21-87\\
\\
Zuckerkandl, E. and L. Pauling 1965. Molecules as documents of evolutionary history. Journal of Theoretical Biology 8: 357-366\\

\section*{Figure Legends}

\pmb{ Figure 1}:\\
\vskip.1in
This set of figures shows plots of the cumulative density functions for the exact solution (Eq. 2.8), the second order approximation (2.9) and the asymptotic
approximation for the case where $\rho $=0.02 and $\omega $=0.05.\\
\\
a. Here $\theta $=0.01, so that the second order approximation converges closely to the exact solution. The lower curve represents the asymptotic
distribution, while the exact solution and second order approximation are nearly superimposed.\\
\\
b. For $\theta $=0.1, the second order approximation remains close to the exact solution. Note that the asymptotic is independent of $\theta $, so
that its divergence from the exact solution is constant in a-c.\\
\\
c. When $\theta $=0.4, the second order approximation deviates significantly from the exact solution, even when the sum is taken to the \(10^6\). This suggests that first and second
order approximations are only robust when the death rate is sufficiently smaller than the birth rate.\\
\\
\\
\pmb{ Figure 2}:\\
\vskip.1in
These figures show cumulative density functions Prob.[$N(t)\leq n \mid \frac{\rho}{\omega}, \theta$] for the same distributions as in Figure 1, but in this instance $\rho $=0.01 and $\omega $=0.1.\\
\\
a. Again, $\theta $=0.01, so the second order approximation is very close. Furthermore, because $\rho $/$\omega $ is fourfold smaller than in Figure
1, the asymptotic approximation is much closer to the exact distribution.\\
\\
b. For $\theta $=0.1, the second order approximation is still only minimally divergent. The asymptotic remains a robust approximation in this instance,
being only dependent on $\rho $/$\omega $.\\
\\
c. When $\theta $=0.4, the same is observed as in Figure 1c.

\section*{Table { }Legends}
\vskip.2in

\pmb{ Table 1}:\\
\vskip.1in
The cumulative probability densities Prob.[$N(t)\leq n \mid \frac{\rho}{\omega}, \theta$] of the exact solution (E), the second order approximation (S), and the asymptotic approximation (A) are shown for n=10, 20,
50, 100, 200, 500, 1000, 2000, and 10000, up to three significant digits. In the first half of the table, the ratio of lineage origination rate to
net birth rate $\rho $/$\omega $=0.25, while $\rho $/$\omega $=0.1 in the second portion. The densities are compared for $\theta $=0.01, 0.1, and
0.4, again showing the breakdown of the second order approximation for large $\theta $, and of the asymptotics for larger $\rho $/$\omega $.\\
\\
\\
\pmb{ Table 2}:\\
\vskip.1in
The critical numbers of individuals $n^*$ corresponding to extremal probability densities p=0.05 and 0.01 are shown for the same parameter values and the same distributions as in Table 1.

\end{document}